\documentclass[a4paper,12pt]{article}
\usepackage{hyperref}
\usepackage[english]{babel}
\usepackage[T1]{fontenc}
\usepackage[utf8]{inputenc}
\usepackage{amsfonts}
\usepackage{amssymb}
\usepackage{amsmath}
\usepackage{graphicx}
\usepackage{cite}
\usepackage{epsfig}
\usepackage{psfrag}
\usepackage{tikz}
\usetikzlibrary{matrix,arrows,decorations.pathmorphing,positioning} 
\usepackage{url}
\usepackage{amsthm}
\usepackage{mathpartir}
\usepackage{enumerate}

\date{}

\title{\bf Pregeometry, Formal Language and Constructivist Foundations of Physics} 

\author{Xerxes D. Arsiwalla$^{1, 4, }$\footnote{\url{x.d.arsiwalla@gmail.com}}  {}    Hatem Elshatlawy$^{2, 3, 4, }$\footnote{\url{hatem@wolfram.com}}
{}    Dean Rickles$^{3, }$\footnote{\url{dean.rickles@sydney.edu.au}}\\  
{}  \\
{\it \small $^{1}$Pompeu Fabra University, Barcelona, Spain}\\ 
{\it \small $^{2}$RWTH Aachen University, Aachen, Germany}\\
{\it \small $^{3}$University of Sydney, Sydney, Australia}\\
{\it \small $^{4}$Wolfram Research, USA}
}

\begin{document}

\maketitle

\begin{abstract}
How does one formalize the structure of structures necessary for the foundations of physics? This work is an attempt at conceptualizing the metaphysics of pregeometric structures, upon which new and existing notions of quantum geometry may find a foundation. We discuss the philosophy of pregeometric structures due to Wheeler, Leibniz as well as modern manifestations in topos theory. We draw attention to evidence suggesting that the framework of formal language, in particular, homotopy type theory, provides the conceptual building blocks for a theory of pregeometry. This work is largely a synthesis of ideas that serve as a precursor for conceptualizing the notion of space in physical theories. In particular, the approach we espouse is based on a constructivist philosophy, wherein ``structureless structures'' are syntactic types realizing formal proofs and programs. Spaces and algebras relevant to physical theories are modeled as type-theoretic routines constructed from compositional rules of a formal language. This offers the remarkable possibility of taxonomizing  distinct notions of geometry using a common theoretical framework. In particular, this perspective addresses the crucial issue of how spatiality may be realized in models that link formal computation to physics, such as the Wolfram model. 
\\
\\
{\bf Keywords:} {Constructivist Foundations of Physics; Formal Language;  Pregeometry;  Topos Theory; Quantum Gravity.}
\end{abstract}

\clearpage

\tableofcontents 

\section{Introduction}

Ever since the inception of general relativity and quantum field theories early and mid-$20^{th}$ century, an outstanding open question in theoretical physics has concerned the quantum nature of gravity, or equivalently, the quantum geometry of space and time at the Planck scale. Contemporary approaches to quantum gravity today have thrown up a rather wide range of proposals on the question of  what the underlying building blocks of quantum geometry may be: from discretizing topology \cite{dowker2006causal,rideout2009emergence}, to triangulated spacetime foam \cite{loll2019quantum}, to geometric operators \cite{rovelli2008loop,oriti2006group},  to extended objects as quantum gravitational fluctuations  \cite{becker2006string}, to holographic duals \cite{aharony2000large}. What perhaps unites these ostensibly diverse theories is the recognition of a pregeometry at the foundations of spacetime, which appears at energies close to the Planck scale. 

How then does one undertake a comparative investigation of pregeometric structures? More specifically, is there a universal structure underlying pregeometric building blocks of physics? This work is an attempt at examining the metaphysics of pregeometric structures. That necessitates a conceptual analysis of the structure of structures upon which existing notions of quantum (or at the least, non-classical) geometry can be founded. Based on tools from formal language theory, this work is a philosophical attempt at addressing  a meta-theory of structures, such that different formulations of  quantum / non-classical geometries may be investigated within a common theoretical framework. It is hoped that the synthesis of ideas presented in this work might pave the way towards a mathematical ``theory of pregeometry,'' which may serve as a formal unifying framework for conceptualizing and analyzing precise definitions of quantum and classical spaces. 

The term ``pregeometry'' was first coined by John A. Wheeler as an  approach to the foundations of physics that ought to   encompass  any underlying explanation of spacetime or quantum gravity (as per Wheeler, this would also include an explanation of elementary particles)   \cite{misner1973gravitation},  \cite{wheeler1980pregeometry}. One may argue that this term merely functions as a placeholder for whatever more elementary structure is eventually found to serve its intended function. Wheeler treated the problem as a kind of exercise in structure-substitution. That is, test every known structure, ``from crystal lattices to standing waves and from Borel sets to the calculus of propositions'' \cite[pp. 17-18]{archibald1977genesis}.   As an historic anecdote, it was also the inadequacy of most seemingly plausible structures that led Wheeler to his ideas encapsulated in the phrase ``It from Bit''. For our purposes here, by pregeometry we will refer to symbolic/linguistic structures which do not come endowed with any pre-assigned geometric attributes. Instead, geometric (and also non-trivial topological) structures should be derived properties of abstract building blocks (within suitable limits, of course). As we will discuss here, higher homotopical constructions in formal languages, expressed using higher categories, turn out to operationalize such a framework of pregeometry. This exercise may be seen as a modern day incarnation of Wheeler's original intuition.

We will argue that a meta-theory of structures, or for that matter any analysis concerning the structure of structures, would be incompatible with a metaphysics based on material realism. That would lead to the well-known infinite regress problem. Even a `Foundationalist' stance with its ``universal self-evident truths'' may not provide a satisfactory resolution to the problem.  Instead, we posit that a `Coherentist' philosophy of physics based on mathematical constructivism provides the appropriate foundations for the kind of pregeometric structures that Wheeler had in mind. Typed languages, and in particular, computational languages, are inherently constructivist. We will argue that what we refer to as ``structureless structures'', are in fact syntactic entities (or types) that realize programs (or proofs). The study of  ``pre-physics'' is then presented as a constructivist paradigm, where spaces and algebras relevant to physical theories are modeled as computational routines built from compositional rules of the underlying formal language.

Apart from Wheeler, language-theoretic approaches to the foundations of physics have also been proposed in several earlier studies pioneered by Isham and collaborators, within the context of topos theory \cite{isham1995structural,isham2000some,doring2008topos1,doring2008topos2,doring2008topos3,doring2008topos4,doring2010thing5}. More recently, the Wolfram Model \cite{Wolfram2020,Arsiwalla2020,arsiwalla2021pregeometric}, Constructor Theory \cite{deutsch2013constructor}, Categorical Quantum Mechanics, \cite{abramsky2009categorical,coecke2018picturing}, Quantum ZX Calculus \cite{Coecke2009a,Gorard2020c,Gorard2021a,Gorard2021b}, Mechanics from Intuitionistic Mathematics   \cite{gisin2021indeterminism}, Operator Mechanics \cite{arsiwalla2022operator}, and Assembly Theory \cite{sharma2023assembly}, can all be seen to be instances of different syntactic formalisms, possibly expressible within a broader language-theoretic framework. Also, some of the modern conceptions in theories of quantum gravity (those mentioned above) such as spin networks, causal sets, group field theories, simplicial calculus, $C^\ast-$algebras, tensor networks, matrix models, and so on, being purely algebraic, can  potentially be constructible within the context of a formal language internal to an appropriate (higher) topos\footnote{This is soon to be reported in a forthcoming publication.}. The key point is that the framework of formal language subsumes many of the above-mentioned theories and  models, and thus provides the appropriate foundation for discussing various notions of pregeometry and geometry proposed in theories of quantum gravity.

The outline of this paper is as follows: To set the motivation for pregeometry as `structureless structure' expressed within a formal language, in Section 2 we begin with a classification of four types of investigations surrounding the foundations of theoretical physics. Based on that, Section 3 leads to a discussion on foundationalism versus coherentism as philosophies of physics. Following the latter, in Section 4 we build the case for pregeometry as structureless structure. Section 5 presents key ideas from  Leibniz's Monadology \cite{leibniz1989monadology}, as well as Pauli-Jung's monism \cite{atmanspacher2022dual}, both of which, one may argue, allude to modern-day ideas of pregeometry. Then in Section 6, we elaborate how formal languages express structureless structures. Finally, in Section 7 we conclude with closing remarks and future directions.

\section{Types of Theories of Fundamental Physics}

By and large, investigations probing the modern-day foundations of theoretical physics\footnote{For our purposes here,  we restrict our discussion to the fundamental physics of particles, fields and geometry. One could  reasonably well build a case in favor of including principles from condensed matter theory. And in fact,  that very may well be the logical extension of the work discussed in this paper.} can broadly be categorized into four main classes (though not completely unrelated to each other): 

\begin{itemize}
\item  (i) Those involving \textbf{interpretation}s of existing physical concepts and structures;  
\item  (ii) Those involving new physical mechanisms or new \textbf{phenomenological models} (often relying on existing theoretical frameworks); 
\item  (iii) Those involving \textbf{new physical structures}, generalizing existing physical notions of space, time or matter in the search for new physics; and  
\item  (iv) Those investigations seeking \textbf{radical new conceptualizations} of existing physics, in order to address questions related to ontological and metaphysical origins of structure and the role of observers.  
\end{itemize}

Efforts involving (i) concern issues such as wave function realism \cite{albert2013wave}, quantum measurement and contextuality \cite{griffiths2019quantum},  the nature of wave-particle duality  \cite{angelo2015wave}, the problem of time in quantum gravity  \cite{anderson2012problem}, and many others.    

\noindent  Examples that fall within class (ii) involve prospective mechanisms to explain the free parameters, such as the masses and couplings of the standard model \cite{froggatt2003trying},  the cosmological constant problem    \cite{weinberg1989cosmological}, the dark matter/modified gravity puzzle  \cite{de2017dark,sanders2002modified}, etc. While investigations of this kind seek to extend known models of particle theory or cosmology, they do so largely within or with limited modifications to the existing structural framework of quantum field theories and general relativity. 

\noindent  On the other hand, present-day efforts involving (iii) typically concern new proposals for quantum gravity \cite{rovelli1998strings,loll2022quantum},  non-perturbative completions of gauge theories \cite{dijkgraaf2002matrix,arsiwalla2006phase,arsiwallasupersymmetric}, noncommutative geometry \cite{connes1994noncommutative}, physics of higher dimensions \cite{randall2006warped},  emergence of spacetime from holography \cite{swingle2018spacetime}, standard model symmetries from division algebras \cite{furey2016standard}, among many others. Investigations of this third type often propose resolutions for outstanding foundational problems by invoking new physical principles and/or new mathematical structures that seek to generalize or extend the scope of  existing frameworks of theoretical physics.   

\noindent And finally, efforts involving class (iv) investigations involve questions regarding the origins of space, time and matter itself. Rather than generalizing existing physical structures, these investigations seek new ontological origins such that existing physical structures may be seen to be emergent or derivable from an underlying ``pre-physical'' framework. The classic example of this is Wheeler's pregeometry    \cite{misner1973gravitation,wheeler1980pregeometry}.

Historically, several of the then new ideas and developments in quantum theory and relativity  originated as new mathematical formalisms of class (iii) and eventually became amenable to investigations of class (ii) and subsequently (i). This was true even for what were considered abstract mathematical structures back in the day. Examples include Clifford algebras and spinors introduced in quantum theory (which eventually made their way into fermionic quantum field theories) by Dirac or non-Euclidean geometries of spacetime by Einstein and Grossmann.  Falsifiable theories, at the very least, are marked by a transition from class (iv) or class (iii) to  class (ii). Based on this premise, the issue isn't whether or not new mathematics is necessary to inform the foundations of physics, but rather, whether or not new mathematical formalisms will eventually become amenable to investigations of class (ii) and class (i).   In particular, investigations seeking to address the origins of established theoretical frameworks such as quantum field theories or cosmological models, will inevitably require the introduction of new mathematical structures, very likely, radical new ones too. The challenge then being: How does one effectively filter out choices that do not refer to the observable universe?

Consider for instance, the longstanding problem of reconciling quantum field theory with general relativity. It is widely believed that such a reconciliation between these founding pillars of modern physics will markedly alter our understanding of major  open  problems in theoretical physics today, including, the origin of our universe, the origin of matter  and  the fundamental forces, the quantum mechanics of black holes, and the nature of space and time; among others. A key question underlying many of these issues is the following: What are the underlying building blocks of space, time and quantum fields? Any theory attempting to bring together  quantum field theory with general relativity will, at the very least, have to take a definitive stance on the nature of these building blocks.  While it seems there is widespread consensus on the view that space, time and matter are fundamentally discrete \cite{isham1995structural}, \cite{gibbs1995small};  specific proposals concerning the nature of this discretization (often described in terms of ``quanta'' of space, time or matter) and consequently the underlying mathematical structure one needs to start with, differ quite a bit. Notable examples include: (a) Theories of quantum gravity, including Loop Quantum Gravity (LQG) \cite{rovelli2008loop},   String Theory and its proposed non-perturbative completion, M-theory  \cite{becker2006string}, Group Field Theory 
\cite{oriti2006group}, the Causal Sets Program  \cite{dowker2006causal},  Causal Dynamical Triangulation (CDT) \cite{loll2019quantum};  (b)  Approaches seeking a unification of the fundamental forces, either within the context of Supersymmetric Gauge  Theories  \cite{altarelli20015}, or within the framework of F-theory  \cite{beasley2009gutsI}, or based on the representation theory of exceptional Lie algebras such as those associated to the group $E_8$   \cite{lisi2010unification};  (c) Models of Emergent Spacetime from entanglement entropy \cite{swingle2018spacetime},  from Energetic Causal Sets \cite{cortes2014universe}, from Emergent Gravity \cite{verlinde2017emergent}, the AdS/CFT correspondence  \cite{aharony2000large}, and other realizations of black hole holography  \cite{arsiwalla2011degenerate,arsiwalla2009entropy,arsiwalla2008more}; (d) Pregeometric models as building blocks of spacetime, as those initiated by Wheeler \cite{misner1973gravitation}, \cite{wheeler1980pregeometry}, and recent approaches based on Homotopy Type Theory \cite{Arsiwalla2020,arsiwalla2021homotopy,arsiwalla2021pregeometric}.   

Extracting empirical verifiability from the multitude of above-mentioned mathematical proposals and filtering out physically redundant ones has been and continues to be a rather daunting challenging. Recent collaborative  projects in quantum gravity phenomenology seek in part to address this problem (see \cite{addazi2022quantum} for a status review of the field). Nonetheless, the point remains that each of these competing proposals of quantum gravity find themselves having to introduce into the foundations of physics new abstract mathematical structures in order to attempt extensions beyond or reconciliations between our existing notions of space, time and matter. This is the essence of class (iii) and class (iv) investigations. Even in a future scenario, where one or more of the above proposals turns out to be an empirically adequate (or at least falsifiable) description of quantum gravity, questions about the origins of that new theory will potentially remain open and require new structures and extensions beyond its existing framework.

\section{The Epistemic Regress Problem: Foundationalism  vs. Coherentism}

Any classification of theories, as the one presented above, into more and more fundamental ones, confronts us with an obvious philosophical problem. For instance, consider the principle of `inferential justification' in the context of epistemology \cite{sep-epistemology}. This states that: ``To be justified in believing A on the basis of B one must be, (1) justified in believing B, and (2) justified in believing that B makes probable A''  \cite{sep-justep-foundational}. Hence, any approach to the foundations of physics based exclusively on inferential justification, in the sense of seeking explanations of explanations recursively, leads to an infinite regress problem.  This is particularly relevant if one were to insist on a fundamentally materialist ontology.  

Of course, issues such as the one stated above, have extensively been discussed in the philosophy of physics, and more generally, in theories of epistemic justification \cite{sep-epistemology}. Typical resolutions of this problem adopt either a stance of `foundationalism' or `coherentism'. 

\begin{itemize}
\item  Foundationalism  assumes the existence of certain self-evident truths that can, in principle,  halt the regress \cite{sep-justep-foundational};
\item Coherentism  requires that statements within that system self-cohere, forming an inter-dependent web of of mutual justification \cite{sep-justep-coherence}. 
\end{itemize}

In the context of physics, the search for realizations of foundationalism  could manifest in terms of what is sometimes referred to as the `final theory'\footnote{See \cite{blum2019heisenberg} for a nice historical account of  Heisenberg's Weltformel (World Formula), a final theory reducing all of physics, known and unknown, to the interactions of one elementary quantum field.}.  On the other hand, proposals grounded in  coherentism  emphasize the role of relations (rather than objects) and compositionality, such that  attributes of a system are described in relational terms.

More specifically, one may ask, what kind of relational frameworks should one examine as plausible candidates for investigating how structural properties of space, time and matter might emerge? Our answer to this is to consider formal languages as the framework for pregeometric foundations of physics.  

\begin{itemize}
  \item  A \textbf{Formal Language} ${\cal L}$ refers to a collection of ``well-formed'' strings $\Sigma^*$ over an alphabet $\Sigma$ (usually a set of finitely many letters), where the superscript "$*$" denotes the Kleene product over $\Sigma$ (the free monoid over the alphabet). The moniker ``well-formed'' refers to syntactic constraints that can be specified either by a generative grammar or a set of $n-ary$ relations based on a universal algebra.  These relations  recursively specify how the letters of $\Sigma$ can be composed to form strings in ${\cal L}$. 

\item  For use in theorem proving or automated reasoning protocols, it is often useful to additionally equip the language ${\cal L}$ with a deductive system.

\item  A \textbf{Deductive System} (also called \textbf{Proof System}) within a language ${\cal L}$ consists of an axiom schema  and / or a collection of inference rules, which can be used for theorem-proving within ${\cal L}$. 

\item  We will call a formal language equipped with a deductive system as a \textbf{Computational Language}. This is sometimes also referred to as a formal system.  

\end{itemize}

As alluded to above, we shall mainly be interested in formal languages that are computational languages. These are typically typed languages equipped with a deductive system, using which one may construct proofs and programs. Computational languages are inherently constructivist in the sense their proof systems only allow constructivist mathematical proofs (those in which proofs by contradiction are not permitted; consequently, a proof that a given property holds, requires constructing an algorithm that realizes an instance of that property). These systems are based on Intuitionistic logic, which as we shall see, allows for expressing theories of physics that are not bound to an \emph{a priori} use of continuum notions. 


In fact, this brand of constructivism founded in formal language is not exclusive to physics. It turns out that many type theories originating from a constructivist paradigm of mathematics are formal systems that are distinct from ZFC set theory (in that, they do not enforce the law of the excluded middle or the axiom of choice). Rather, many of these type theories and their associated  toposes based on the univalence axiom of homotopy type theory \cite{Program2013,Ahrens2021}. In particular, this suggests that there does not exist  merely one  preferred axiomatization to describe all of mathematics, as proponents of the Hilbert school of thought may have hoped for.  Rather, there are multiple universes (or toposes) where mathematics can be formulated. These universes may be founded on very distinct axiomatization schemes, which nonetheless can be transformed from one to another \cite{Shulman2016}, \cite{Shulman2017}.  

How does formal language relate to pregeometry? In the next few sections we will argue that our  ``structureless structures'' are in fact syntactic entities (types) that realize programs (proofs) in a formal language. Within the modern set-up of Homotopy Type Theory, spaces and algebras relevant to physical theories can be modeled as computational routines built from compositional rules of a formal language.  Arguably, the kind of constructivism guiding the current foundations of mathematics turn out to be important for the foundations of physics too.

\section{Pregeometry as Structureless Structure}

What kind of structures should a constructivist coherentism entail for class (iv) investigations relevant to the foundations of physics?  As alluded to above, formal language, and in particular, computational languages based on syntactic structures, their compositions and rules of computation provide the constructivist building blocks for a realization of pregeometry that does not hinge on pre-existing notions of space, time and matter. Such formal constructs necessarily shift away from a materialist ontology of physics. 

\begin{itemize}
  \item   \textbf{Structureless Structures} thus refer to symbolic and relational entities of a formal language.  These symbolic structures are the building blocks of proofs, programs and computations.   

\end{itemize}

When Wheeler introduced the notion of pregeometry, he thought about it as an all-encompassing approach to the very foundations of physics, with the idea that pregeometry ought to transcend any structural explanation of space, time,  matter and even physical law  \cite{misner1973gravitation,wheeler1980pregeometry}. Pregeometry was thus intended as a broad conceptual framework from which one may seek, or upon which one may build, descriptions of quantum gravity. As mentioned, at the time, Wheeler treated the problem as a structure-substitution exercise, meaning that he tested every known structure, with the objective of seeking structural abstractions that might serve as the building blocks of the physical universe. In particular, he examined abstractions of lattices, waves, Borel sets and importantly, the calculus of propositions. Additionally, Wheeler also introduced what he referred to as  ``Observership''. This he deemed as crucial for any physical theory \cite{archibald1977genesis} (see \cite{blum2016experience} for a recent discussion on Wheeler's ideas of observership and their relation to our experiences of space and time). Indeed, some of these ideas eventually led him to his now well-known `mantra': ``It from Bit''.

Yet another angle that inspired Wheeler to considering pregeometry as the foundation of all of physics came from a gravitational collapse argument (``the crisis of collapse''): if the universe can collapse, then it will take space, time, matter, and law with it;  therefore, there needs to be something that transcends these in such a way as to not be subject to the same demise (see \cite{blum2022john} for an insightful historical overview). Moreover, this something (the pregeometry) needs to be such that it can forge a way for the universe to come into being. It must provide building blocks. Hence, it serves cosmogonical and cosmological functions in Wheeler's thinking. This crisis of collapse curtails the kind of approach Einstein was attempting (which Wheeler also initially pursued in his earlier geometrodynamical investigations), in which space itself is the primordial substance from which all else is constructed. Despite the impressive topological gymnastics involved in constructing mass, charge and (with extreme difficulty) spin from space alone, contemporary theory is simply inadequate to the task when the collapse problem is faced.

The sum-over-histories ideas were part of an initial simple attempt to quantize gravity using what would now be recognized as integrating over moduli spaces of geometries and fields. It was largely Wheeler's PhD student Charles Misner who did this early work on quantum gravity \cite{misner1973gravitation}. The aim was to try and generate more structure in space by allowing for quantum fluctuations of the geometrical (and topological) properties, in order to produce multiply-connected wormholes that could be used to thread fields and explain how point-charges can appear to emerge in a purely continuous field theory.  However, in addition to the collapse problem, Wheeler was also motivated by the a desire for the deeper constructibility  of the world of physics. As he put it using an analogy:

\begin{quotation}
\noindent Glass comes out of the rolling mill looking like a beautifully transparent and homogeneous elastic substance. Yet we know that elasticity is not the correct description of reality at the microscopic level. Riemannian geometry likewise provides a beautiful vision of reality; but ... is inadequate to serve as primordial building material. \cite[p. 544]{patton1975physics}
\end{quotation}

\noindent Central to this new approach to physics, in which one seeks the deepest level of structure, is that idea that one ought not to start from the upper levels in order to figure things out. In other words, part and parcel of Wheeler's approach was a quite radical constructivism. It is no good, from this point of view, to consider conventional quantization approaches in which one \emph{begins} with the classical system and then applies a procedure to it. This corresponds to our artificial methodology, rather than nature's own technique for creating the world which is, after all, already quantum. What pregeometry and constructivism share, and what we also share, is the belief that our aim, in foundational work, must be to find the methods and materials that nature herself uses to build the world.

It is worth remarking that that many of the contemporary theories of quantum gravity, in fact, are set up with a fair share of \emph{a priori} geometric structures  (see examples (a) - (c) mentioned earlier). On the other hand, a truly pregeometric  description of the kind Wheeler had in mind, ought to be one from which all geometric features of the physical universe should be derived  (\cite{meschini2005geometry} discusses this point at length).  Hence,  it is the precursors of geometry (and one may argue, even topology) that  make up the genuinely pregeometric building blocks of the universe.  Now, given the common expectation, that blending general relativity and quantum mechanics,  would permit ``foam-like'' realizations of geometry, at energies close to those that existed in the early universe, this implies, at the very least, that a set of more fundamental rules regarding connectivity of spacetime that are independent of topology and dimensionality, are required (as emphasized by Wheeler).  Formulations of theoretical physics based on  pregeometric structures then allow one to work with deeper underlying rules   that are not dependent on classical structural assumptions about the properties of space and time.  As we shall see, such an approach capitalizes upon deep connections between theoretical physics, computation, proof theory and homotopy. It is such pregeometric entities that we will hereon in refer to as ``Structureless Structures''\footnote{Here we arrive at this with a focus on pregeometry. However, connections between physics, computation and formal systems have been discussed in other contexts too. For instance, relating to undecidable dynamics and the edge of chaos in  \cite{prokopenko2019self}; or founded on monoidal categories, quantum processes and cobordisms  in \cite{baez2011physics}.}.

\section{Pregeometry in Metaphysics }  

Of course, the idea that "something" has to transcend space, time and matter has long since been part of philosophical discourse, in particular, metaphysics \cite{risi2007geometry}. One of the early proponents of relationalism in the metaphysics of space and time, was none other than Gottfried Wilhelm Leibniz (in contrast to his contemporary at the time, Isaac Newton). Leibniz, however, was thinking of something beyond merely material or physical relationalism. Those views led him to the concept of `monads' \cite{leibniz1989monadology}. Leibniz's \emph{Monadology} was an  attempt to codify an entire philosophical system. For all its encompassing majesty, it was notable for its extreme brevity\footnote{In fact, it was so brief that Leibniz later added annotations pointing the reader to other works for clarification.}. The Monadology can in all likelihood also be viewed as the first example of a pre-space theory. Leibniz argued in favor of a set of features of space from principles applying to a set of relations that are not spatial themselves. That is, the relations between monads are used to set up a correspondence to phenomenal space, with its characteristic features such as extension\footnote{See \cite{risi2007geometry} for a detailed study of Leibniz's deeper philosophy of space.}. In this sense, his monadological theory of space is more primitive than the usual kind of physical relationalism, in which relations are simply thought to involve the objects of physics, with spatial relations being secondary to those objects (i.e. supervenient) in an ontological sense\footnote{Wheeler himself was influenced by some of Leibniz's ideas on space, time and matter. A historical account of this intersection of ideas can be found here: \cite{furlan2020merging}.}. In Monadology, a different kind of object ends up being primary (i.e. the monad is fundamental), and the relations hold between these such that both space, the objects of physics, and every other thing in the manifest world, emerge from this more basic layer. Moreover, monads are \emph{simples} (``the true atoms of nature'') in the sense of admitting no further reduction or decomposition into other elements. That is, they have no structural elements of their own and so provide a kind of structureless structure. And yet, from this simple foundation, according to Leibniz, we can generate all of the incredible complexity of the world.

The monads collectively provide all possible perspectives of a world, as tiny independent mirrors (or points of view). However, there is also a sense in which the monads are carrying out a pre-set program (or entelechy), coordinated with all other monads, in a pre-established and divinely choreographed dance determined to generate (i.e. construct) the best of all possible worlds. While there are the well-known principles of sufficient reason and identity of indiscernibles providing basic constraints on this construction, the principles themselves do not directly determine what is constructed. Rather, they inform the composition of the monads into complex structures which is then carried out through the pre-established harmony. A major reason for the introduction of  pre-established harmony was to explain the mind-body (or soul-body) correlations. For Leibniz there was no causal link and the correlation simply follows from the common cause in which both were set on their way. 

In this context, it is interesting to note the philosophical parallels of Leibniz's metaphysics   with Wolfram's model of physics (or rather, ``pre-physics'') \cite{Wolfram2020}. Analogous to Leibniz's monads, in the Wolfram model, abstract rewriting events comprise the ``atoms of nature''. These events are generated by rewriting rules, that realize abstract computation. Based on local rule application, rewriting events are knitted together via causal relations. These are the causal graphs of the Wolfram model. The irreducible rewriting events and their mutual relations, taken in some appropriate limit, are hypothesized as models of spacetime geometries \cite{Arsiwalla2020,arsiwalla2021homotopy,arsiwalla2021pregeometric}. Furthermore, Wolfram's model has a remarkably similar explanation for the correspondence of the world to the mind in that they both emerge from the same initial rules for construction and emerge in parallel with the mind (or observer) simply sampling the world and providing a perspective \cite{wolframruliad,wolframconc}, much like a monad, where different observers perceive the whole universe from different points of view. Likewise, one can find a similar generation of variety in the Wolfram model through this dislocation of a single, unified structure into many of points of view  \cite{elshatlawy2023ruliology}.

Of course, Leibniz's theory, as it stands, cannot provide a satisfactory foundation for physics. At least, not one of much practical value in terms of showing how our present theories and phenomena can be \emph{constructed}. Our aim in this work is to discuss some developments, including very recent ones, in this direction. Ultimately, the approach we focus on, that is, structureless structure from formal language, places the ontological weight on the very rules of construction themselves. By contrast with Leibniz's ``God as architect'' (as he puts it in S.89 of his \emph{Monadology}), here the metaphor is better expressed as ``Nature constructing itself", in particular, space, time, matter and law. 

Besides Leibniz, notions of pre-physical substrates of existence have also been discussed in the philosophy of mind, in particular, ideas related to monism of mind and matter \cite{atmanspacher2022dual}. In the context of the mind-body problem, the philosophy of monism seeks to resolve the metaphysical  debate between physicalism and idealism by proposing a fundamentally new neutral substrate that is by itself neither physical nor mental, but instead, whose various manifestations then realize the physical and mental components of the world. A prominent example of this school of thought is dual-aspect monism proposed by the physicist Wolfgang Pauli and the psychologist Carl Jung. Dual-aspect monism posits that the physical and mental are merely complementary perspectives of an underlying neutral substrate \cite{atmanspacher2022dual}.  In other words, the physical universe, including mental states of agents within it, are to be built upon a metaphysically fundamental layer of reality that is both pre-physical and pre-mental. From this perspective, monism necessitates pregeometric building blocks for our perceived reality.

\section{Pregeometric Theories from Formal Language }

What then should constitute the essence of pregeometric structure from which space, time and matter all emerge? For one, pregeometry being structureless structure, cannot arise from yet another unbeknownst physical substrate. A foundationalist philosophy of pregeometry can potentially admit an irreducible underlying substrate. But then, one would need to posit that the existence of such a substrate be accepted as a ``universal self-evident truth'', such that one cannot ask further questions about its nature or origins. This may seem a rather unsettling predicament to have to accept within a scientific theory. Furthermore, even if such a universal self-evident entity existed, how should one describe it (as opposed to explaining it) within a given theory without alluding to any spatial or temporal notions, including internal spaces (those describing internal degrees of freedom corresponding to internal symmetries, spin or gauge indices)? On the other hand, a coherentist philosophical stance places precedence on relations rather than objects and posits that structure emerges from the metaphysics of abstract relations. In a philosophy of this kind, emphasis is placed on the ontology of relations rather than the ontology of objects.

It turns out that the appropriate mathematical framework to formalize a theory of abstract relations and their properties is what is called a `Formal System' with well-defined rules of compositionality. Formal languages are precisely such systems. Formal systems lie at the heart of mathematical logic, computer science, cryptography and several other branches of mathematics. More pertinently, formal language, and in particular,  Homotopy Type Theory and the Univalent Foundations program have been at the forefront of important recent advances seeking a new constructivist foundation for all of mathematics \cite{Program2013,Shulman2016,Shulman2017}. Here, we seek to identify appropriate parallels arising from developments in mathematics to the foundations of physics. We reckon that the application of homotopy type theory and its representation in higher category theory will be extremely useful for:  \\ 
(i) Exploring higher symmetries and spaces in physics, that cannot readily be captured by current methods; and for \\ 
(ii)  Seeking a constructivist foundation for physics, where structures intrinsic to notions of the continuum are not fundamental, but emerge within well-defined limits.  

While Wheeler himself had suggested a propositional calculus as a pregeometric framework from which the emergence of physical structures may be sought (though he ultimately had to introduce a `participator' to deal with undecidable propositions of physics); our contention here is that formal language permits the expression of generic pregeometric calculi. As mentioned, this parallels the way mathematicians discuss universes of mathematics using Homotopy Type theory. A formal language encompasses a system of primitive symbols (or ground types) along with relations for constructing composite types which can be used to construct clauses and sentences. The latter constitute propositions expressible within the language. A language can  additionally be equipped with axioms and inference rules for a deductive logic using which one can reason about its propositions. However, propositions are only declarative statements. One can go further. Including variables and quantifiers allows one to extend a propositional system to one that includes predicates, thus expressing formulae, whose validity (truth) may subsequently be evaluated within a specified interpretation (semantic modality). Based on the logical relations and inference rules that a given language admits, one can then construct proofs relating one formula to another, that is, prove theorems within that language. 

Besides Wheeler's pregeometry, the role of formal language towards conceptualizing new foundations for quantum theory and physics on discrete spaces has  been extensively investigated by Chris Isham and collaborators \cite{isham2000some,doring2010thing5}. Rather than pregeometry per se, the motivations for the latter arose in seeking an axiomatization of physical theories within a common mathematical framework - that of topos theory. Toposes are categories that behave like \textbf{Sets} (the category of sets). Like \textbf{Sets}, toposes are equipped with the category-theoretic analogue of Cartesian products, disjoint unions, a singleton set, a notion of a set of functions, and importantly, a notion of sub-sets of objects (i.e., sub-objects). Thus, toposes are formal ``places'' where foundations of mathematics can be formulated. Examples of toposes other than \textbf{Sets} are the category of finite graphs, the category of $G$-sets, the category of presheaves over a small category. 

Even more generally, attempts seeking a formal axiomatic framework for physical theories, pre-date Wheeler, going all the way back to David Hilbert. In his 1900 address at the International Congress of Mathematics in Paris, Hilbert stated his famous 23 open problems of mathematics. Of these, the $6^{th}$ problem referred to a universal axiomatization of physics (see \cite{corry1997david} for a historical overview). Apart from the issue of whether such axiomatizations ought to be universal or even complete, it set the course for seeking mathematical formulations of physical theories using a common (axiomatic and / or inferential) framework. Then, towards the latter half of the $20^{th}$ century, with rapid advances in category theory, William Lawvere, one of the founders of categorical logic, sought to build the foundations of mathematics in topos theory (as opposed to set theory) \cite{lawvere1966category}. Lawvere was also interested in applications of topos theory towards the formalization of physics  \cite{lawvere1997toposes}, which was subsequently followed up by Isham and others (as noted above). 

It is worth noting that, while the topos-theoretic foundations discussed here offer the elegant possibility of expressing theories of physics through a ``mathematically unified'' framework, they do not carry the usual baggage of grand unification of physical theories. This allows for a potentially background independent formulation of a broad class of physical theories. In particular, Isham's work proposes specifically distinct toposes for classical and quantum mechanics \cite{doring2008topos1,doring2008topos2}. The key objective  of their program was to do away with any a priori use of continuous spatial or temporal constructs in formulating notions related to classical or quantum systems. As stated in \cite{doring2008topos1}, ``the use of continuous properties associated to space and time would be deemed a major error if those turned out to be fundamentally incompatible with what is needed for a theory of quantum gravity''. The contention there was that theories of a physical system should be formulated within a topos that depends on both, the theory-type and the system-type. In turn, any topos-theoretic approach employs formal language. This is because of a  well known result in topos theory that there exists an internal formal language associated to each topos \cite{johnstone2014topos}. In fact, not only does each topos generate an internal language, but, conversely, a language satisfying appropriate conditions generates a topos \cite{johnstone2014topos}. The goal in \cite{doring2010thing5} was to find a novel structural frameworks within which new types of theory can be constructed, and in which continuum quantities play no fundamental role. These works proposed an abstract language-theoretic formulation of classical and quantum mechanics which primarily addresses questions related to kinematics of classical and quantum systems in arbitrary spaces. Going beyond this kinematical description, the question is how does one generalize topos-theoretic approaches to address pregeometric theories as well as other effective theories at high energies? 

More generally, the kind of languages admissible in toposes are typed languages. Type theory provides the building blocks to formally construct such languages. Given their constructivist flavor, the logic expressed by type theories in toposes is intuitionistic logic. This means one need not enforce the law of the excluded middle or the axiom of choice in these formal systems. Furthermore, the natural extension of intuitionistic type theory is Homotopy Type Theory (HoTT), which includes homotopy $n$-types, up to $\infty$-types. The representation of homotopy types takes us beyond the realm of standard category theory to higher category theory, which includes morphisms between morphism (representing homotopies between types). This tower of higher morphisms goes all the way to $\infty$-categories. Formal languages expressed in homotopy type theory are internalized in higher categories, and consequently higher toposes.  

How then do topological and geometric spaces relevant to physics (and mathematics) arise from type-theoretic building blocks?  One of the key takeaways from the synthetic geometry and homotopy type theory program is that the notion of space arises from functorial constructions involving $\infty$-toposes   \cite{Schreiber2012,Schreiber2013a,Shulman2017}. Geometry is thus 
inherited from higher structures, and induced upon local structures by taking sections or projections of the total space  \cite{arsiwalla2021homotopy,arsiwalla2021pregeometric}.  Homotopy type theory provides a syntactic formalism for realizing higher structures. The objects of the $\infty$-topos under consideration are the so-called `$\infty$-groupoids'. The latter are categories endowed with a tower of higher morphisms, up to infinity (and invertibility conditions). Via Grothendieck's hypothesis, $\infty$-groupoids realize models of  formal topological spaces \cite{baez2007homotopy}.  With additional `cohesivity conditions', one also obtains synthetic geometric spaces in $\infty$-toposes from this construction 
\cite{Schreiber2012}, \cite{Schreiber2013a}, \cite{schreiber2016higher}, \cite{Shulman2016},  \cite{Shulman2017}.  These authors also show how quantum field theories with higher gauge symmetries can be formalized in  $\infty$-toposes   \cite{Schreiber2012}, \cite{Schreiber2013a}, \cite{schreiber2016higher}. Higher homotopical structures in formal languages expressed using higher categories thus provide us a useful formal framework for constructing pregeometric physics as well as theories of higher symmetries. 

Furthermore, a computational realization of the above $\infty$-groupoid constructions was shown in \cite{arsiwalla2021homotopy,arsiwalla2021pregeometric,Arsiwalla2020}.  This construction was based on what are called `Multiway Systems', the non-deterministic rewriting systems of the Wolfram model \cite{Wolfram2002a,Wolfram2020}.  Using a type-theoretic representation of  multiway rewriting systems, \cite{arsiwalla2021pregeometric} provide an algorithmic construction of  higher homotopies on non-deterministic rewriting systems. This connection between abstract rewriting systems and higher homotopies suggests a way to realize spatial structures and geometry from purely pregeometric models such as those based on rewriting systems (mentioned above).

\section{Outlook and Discussions}

In conclusion, this work serves as an initial metaphysical exploration of a plausible description of pregeometric building blocks for the physics of spacetime, based on formal language. We have put forth the proposal that syntactic structures formalized in computational languages model the kind of pregeometric structures that Wheeler had in mind concerning the foundations of physics. We described these  pregeometric structures as structureless structure to emphasize the necessity to shift away from a fundamentally material ontology. Instead, these are symbolic and relational structures of a formal language.  

Our approach to pregeometry takes seriously a constructivist stance on the laws and structures of the physical universe; not merely in terms of how observers may perceive the universe, but more importantly, in metaphysical terms, as to how these laws and structures might come into being. Such a perspective closely aligns with Wheeler's intuitions of pregeometry as something that transcends space, time, matter and law. Indeed, this flavor of constructivism, not surprisingly, resembles the kind of constructivism that has been recognized in recent advances in the foundations of mathematics, particularly in the context of  Homotopy Type Theory and the Univalence Foundations program \cite{Program2013}. Speaking of the ``unreasonable effectiveness of mathematics'' point of view   \cite{wigner1960unreasonable}, it is perhaps fitting that the metaphysics of spacetime geometry directly draws from formal advances in metamathematics. A computational realization of this connection between metamathematics and physics in terms of rewriting systems can be found in \cite{wolfram2022metamathematics}. All in all, a language-theoretic constructivist framing serves as an important conceptual advance for approaches that emphasize the interplay between computation and physics, such as the Wolfram model.

Given that the coherentist constructivism discussed here follows from requiring to go beyond a materialist  ontology for pregeometry, this implies that primitives of pregeometry are not merely discretizations of classical structures in physics. Its about nature constructing itself from abstract computation using syntactic compositions and relations. Simply replacing classical and quantum systems on continuous spaces with their discrete counterparts is unlikely to capture the full essence of Planck scale physics\footnote{It is well-known that quantization itself isn't solely about discretizing a system, or about replacing a system of equations in classical analysis with those in discrete analysis.}.  The plausible emergence of space, time and matter, and theories describing them at or below the Planck length, will likely require new mathematical formalisms that go beyond mere replacements of classical real or complex analysis with discrete geometry. Recent advances in homotopy theory, higher algebra and topos theory offer new mathematical methods for such investigations  (see \cite{lurie2009higher,Riehl2017a,riehl2018elements} for works in this direction). Also, worth noting that the strict dichotomy between continuous versus discrete geometry may be a bit misleading given that there exist examples of geometric formalisms that are by themselves neither continuous nor discrete, such as operator algebras that realize ``pointless geometry'' \cite{connes1994noncommutative,arsiwalla2022operator}. Rather, it is the representation and spectrum of these operators that may take continuous or discrete values under different conditions. This feature has been exploited in models of quantum gravity such as loop quantum gravity, noncommutative geometry and group field theories \cite{rovelli2008loop,connes1994noncommutative,oriti2006group}.  

As mentioned earlier, approaches advocating the use of formal language to conceptualize the foundations of physics, by themselves, are not new. However, the newly developing mathematical formalism of homotopy type theory \cite{Program2013}, extended topological field theories \cite{lurie2009classification}, operator mechanics \cite{arsiwalla2022operator,chester2023covariant}, infinity-categories \cite{riehl2018elements}, infinity-toposes \cite{Schreiber2013a}, higher-arity algebras \cite{zapata2022invitation,zapata2022heaps,zapata2023beyond}, etc offer new ways to investigate pregeometric structures formalized in computational languages. A recurring theme in these investigations is that of higher structures. A language-theoretic pregeometric formalism based on higher structures will likely bridge, or at the very least, help identify crucial intersections between existing constructivist and background-independent approaches to quantum gravity.

\section*{Acknowledgments}

We thank Stephen Wolfram for his enthusiasm, encouragement and discussions on the manuscript. We also like to thank Alex Blum for many useful suggestions and comments on this work.

This research has been supported by the following grants:  The Foundational Questions Institute and Fetzer-Franklin Fund, a donor advised fund of Silicon Valley Community Foundation [FQXi-RFP-1817]; The John Templeton Foundation [Grant ID 62106]; and the Australian Research Council [Grant DP210100919].

\bibliographystyle{splncs04}
\bibliography{hottrefs}

\begin{thebibliography}{10}
\providecommand{\url}[1]{\texttt{#1}}
\providecommand{\urlprefix}{URL }
\providecommand{\doi}[1]{https://doi.org/#1}

\bibitem{abramsky2009categorical}
Abramsky, S., Coecke, B.: Categorical quantum mechanics. Handbook of quantum
  logic and quantum structures  \textbf{2},  261--325 (2009)

\bibitem{addazi2022quantum}
Addazi, A., Alvarez-Muniz, J., Batista, R.A., Amelino-Camelia, G., Antonelli,
  V., Arzano, M., Asorey, M., Atteia, J.L., Bahamonde, S., Bajardi, F., et~al.:
  Quantum gravity phenomenology at the dawn of the multi-messenger era—a
  review. Progress in Particle and Nuclear Physics  \textbf{125},  103948
  (2022)

\bibitem{aharony2000large}
Aharony, O., Gubser, S.S., Maldacena, J., Ooguri, H., Oz, Y.: Large n field
  theories, string theory and gravity. Physics Reports  \textbf{323}(3-4),
  183--386 (2000)

\bibitem{Ahrens2021}
Ahrens, B., North, P.R., Shulman, M., Tsementzis, D.: The univalence principle
  (2021), \url{https://arxiv.org/abs/2102.06275}

\bibitem{albert2013wave}
Albert, D.Z.: Wave function realism. The wave function: Essays on the
  metaphysics of quantum mechanics pp. 52--57 (2013)

\bibitem{altarelli20015}
Altarelli, G., Feruglio, F.: Su (5) grand unification in extra dimensions and
  proton decay. Physics Letters B  \textbf{511}(2-4),  257--264 (2001)

\bibitem{anderson2012problem}
Anderson, E.: Problem of time in quantum gravity. Annalen der Physik
  \textbf{524}(12),  757--786 (2012)

\bibitem{angelo2015wave}
Angelo, R., Ribeiro, A.: Wave--particle duality: An information-based approach.
  Foundations of Physics  \textbf{45},  1407--1420 (2015)

\bibitem{archibald1977genesis}
Archibald~Wheeler, J.: Genesis and observership. In: Foundational Problems in
  the Special Sciences: Part Two of the Proceedings of the Fifth International
  Congress of Logic, Methodology and Philosophy of Science, London, Ontario,
  Canada-1975. pp. 3--33. Springer (1977)

\bibitem{arsiwalla2008more}
Arsiwalla, X.D.: More rings to rule them all: fragmentation, 4d
  $\leftrightarrow$ 5d and split-spectral flows. Journal of High Energy Physics
   \textbf{2008}(02), ~066 (2008)

\bibitem{arsiwalla2009entropy}
Arsiwalla, X.D.: Entropy functions with 5d chern-simons terms. Journal of High
  Energy Physics  \textbf{2009}(09), ~059 (2009)

\bibitem{arsiwallasupersymmetric}
Arsiwalla, X.D.: Supersymmetric black holes as probes of quantum gravity. PhD
  Thesis, University of Amsterdam  (2010),
  \url{https://pure.uva.nl/ws/files/871677/75420_thesis.pdf}

\bibitem{Arsiwalla2020}
Arsiwalla, X.D.: Homotopic foundations of wolfram models  (2020),
  \url{https://community.wolfram.com/groups/-/m/t/2032113}

\bibitem{arsiwalla2006phase}
Arsiwalla, X.D., Boels, R., Marino, M., Sinkovics, A.: Phase transitions in
  q-deformed 2d yang-mills theory and topological strings. Physical Review D
  \textbf{73}(2),  026005 (2006)

\bibitem{arsiwalla2011degenerate}
Arsiwalla, X.D., de~Boer, J., Papadodimas, K., Verlinde, E.: Degenerate stars
  and gravitational collapse in ads/cft. Journal of High Energy Physics
  \textbf{2011}(1),  1--66 (2011)

\bibitem{arsiwalla2022operator}
Arsiwalla, X.D., Chester, D., Kauffman, L.H.: On the operator origins of
  classical and quantum wave functions. arXiv preprint arXiv:2211.01838  (2022)

\bibitem{arsiwalla2021pregeometric}
Arsiwalla, X.D., Gorard, J.: Pregeometric spaces from wolfram model rewriting
  systems as homotopy types. arXiv preprint arXiv:2111.03460  (2021)

\bibitem{arsiwalla2021homotopy}
Arsiwalla, X.D., Gorard, J., Elshatlawy, H.: Homotopies in multiway
  (non-deterministic) rewriting systems as $n$-fold categories. arXiv preprint
  arXiv:2105.10822  (2021)

\bibitem{atmanspacher2022dual}
Atmanspacher, H., Rickles, D.: Dual-aspect monism and the deep structure of
  meaning. Routledge (2022)

\bibitem{baez2011physics}
Baez, J., Stay, M.: Physics, topology, logic and computation: A rosetta stone.
  New Structures for Physics  \textbf{813}, ~95 (2011)

\bibitem{baez2007homotopy}
Baez, J.: The homotopy hypothesis. Fields Institute. Available online at
  http://math. ecr. edu/home/baez/homotopy  (2007)

\bibitem{beasley2009gutsI}
Beasley, C., Heckman, J.J., Vafa, C.: Guts and exceptional branes in f-theory
  -- i. Journal of High Energy Physics  \textbf{2009}(01), ~058 (2009)

\bibitem{becker2006string}
Becker, K., Becker, M., Schwarz, J.H.: String theory and M-theory: A modern
  introduction. Cambridge university press (2006)

\bibitem{blum2022john}
Blum, A., Furlan, S.: How john wheeler lost his faith in the law. In:
  Rethinking the Concept of Law of Nature: Natural Order in the Light of
  Contemporary Science, pp. 283--322. Springer (2022)

\bibitem{blum2016experience}
Blum, A.S., Renn, J., Schemmel, M.: Experience and representation in modern
  physics: The reshaping of space  (2016)

\bibitem{blum2019heisenberg}
Blum, A.S.: Heisenberg's 1958 Weltformel and the Roots of Post-empirical
  Physics. Springer (2019)

\bibitem{chester2023covariant}
Chester, D., Arsiwalla, X.D., Kauffman, L., Planat, M., Irwin, K.: Quantization
  of a new canonical, covariant, and symplectic hamiltonian density. arXiv
  preprint arXiv:2305.08864  (2023)

\bibitem{Coecke2009a}
Coecke, B., Duncan, R.: Interacting quantum observables: Categorical algebra
  and diagrammatics. New J. Phys. 13 (2011) 043016  (Jun 2009).
  \doi{10.1088/1367-2630/13/4/043016}

\bibitem{coecke2018picturing}
Coecke, B., Kissinger, A.: Picturing quantum processes. In: International
  Conference on Theory and Application of Diagrams. pp. 28--31. Springer (2018)

\bibitem{connes1994noncommutative}
Connes, A.: Noncommutative geometry. Springer

\bibitem{corry1997david}
Corry, L.: David hilbert and the axiomatization of physics (1894--1905).
  Archive for history of exact sciences  \textbf{51},  83--198 (1997)

\bibitem{cortes2014universe}
Cortes, M., Smolin, L.: The universe as a process of unique events. Physical
  Review D  \textbf{90}(8),  084007 (2014)

\bibitem{de2017dark}
De~Swart, J., Bertone, G., van Dongen, J.: How dark matter came to matter.
  Nature Astronomy  \textbf{1}(3), ~0059 (2017)

\bibitem{deutsch2013constructor}
Deutsch, D.: Constructor theory. Synthese  \textbf{190}(18),  4331--4359 (2013)

\bibitem{dijkgraaf2002matrix}
Dijkgraaf, R., Vafa, C.: Matrix models, topological strings, and supersymmetric
  gauge theories. Nuclear Physics B  \textbf{644}(1-2),  3--20 (2002)

\bibitem{doring2008topos1}
D{\"o}ring, A., Isham, C.J.: A topos foundation for theories of physics: {I}.
  {Formal} languages for physics. Journal of Mathematical Physics
  \textbf{49}(5) (2008)

\bibitem{doring2008topos2}
D{\"o}ring, A., Isham, C.J.: A topos foundation for theories of physics: {II}.
  {Daseinisation} and the liberation of quantum theory. Journal of Mathematical
  Physics  \textbf{49}(5) (2008)

\bibitem{doring2008topos3}
D{\"o}ring, A., Isham, C.J.: A topos foundation for theories of physics: {III}.
  {The} representation of physical quantities with arrows $\delta^{o}$ $({A})$
  : ${\Sigma}$ $\rightarrow$ ${R}^{\geq}$. Journal of Mathematical Physics
  \textbf{49}(5) (2008)

\bibitem{doring2008topos4}
D{\"o}ring, A., Isham, C.J.: A topos foundation for theories of physics: {IV}.
  {Categories} of systems. Journal of Mathematical Physics  \textbf{49}(5)
  (2008)

\bibitem{doring2010thing5}
D{\"o}ring, A., Isham, C.J.: ``{What} is a thing?'': Topos theory in the
  foundations of physics. In: New structures for physics, pp. 753--937.
  Springer (2010)

\bibitem{dowker2006causal}
Dowker, F.: Causal sets as discrete spacetime. Contemporary Physics
  \textbf{47}(1), ~1--9 (2006)

\bibitem{froggatt2003trying}
Froggatt, C., Nielsen, H.: Trying to understand the standard model parameters.
  Surveys in High Energy Physics  \textbf{18}(1-4),  55--75 (2003)

\bibitem{furey2016standard}
Furey, C.: Standard model physics from an algebra? arXiv preprint
  arXiv:1611.09182  (2016)

\bibitem{furlan2020merging}
Furlan, S.: Merging labyrinths. Studia Leibnitiana (H. 1/2),  123--155 (2020)

\bibitem{gibbs1995small}
Gibbs, P.E.: The small scale structure of space-time: a bibliographical review.
  arXiv preprint hep-th/9506171  (1995)

\bibitem{gisin2021indeterminism}
Gisin, N.: Indeterminism in physics and intuitionistic mathematics. Synthese
  \textbf{199}(5-6),  13345--13371 (2021)

\bibitem{Gorard2020c}
Gorard, J., Namuduri, M., Arsiwalla, X.D.: {ZX}-calculus and extended
  hypergraph rewriting systems {I}: A multiway approach to categorical quantum
  information theory. arXiv preprint arXiv:2010.02752  (2020)

\bibitem{Gorard2021b}
Gorard, J., Namuduri, M., Arsiwalla, X.D.: Fast automated reasoning over string
  diagrams using multiway causal structure  (May 2021),
  \url{https://arxiv.org/abs/2105.04057}

\bibitem{Gorard2021a}
Gorard, J., Namuduri, M., Arsiwalla, X.D.: {ZX}-calculus and extended wolfram
  model systems {II}: Fast diagrammatic reasoning with an application to
  quantum circuit simplification. arXiv preprint arXiv:2103.15820  (2021)

\bibitem{griffiths2019quantum}
Griffiths, R.B.: Quantum measurements and contextuality. Philosophical
  Transactions of the Royal Society A  \textbf{377}(2157),  20190033 (2019)

\bibitem{sep-justep-foundational}
Hasan, A., Fumerton, R.: {Foundationalist Theories of Epistemic Justification}.
  In: Zalta, E.N., Nodelman, U. (eds.) The {Stanford} Encyclopedia of
  Philosophy. Metaphysics Research Lab, Stanford University, {F}all 2022 edn.
  (2022)

\bibitem{isham1995structural}
Isham, C.: Structural issues in quantum gravity. arXiv preprint gr-qc/9510063
  (1995)

\bibitem{isham2000some}
Isham, C.J., Butterfield, J.: Some possible roles for topos theory in quantum
  theory and quantum gravity. Foundations of physics  \textbf{30}(10),
  1707--1735 (2000)

\bibitem{johnstone2014topos}
Johnstone, P.T.: Topos theory. Courier Corporation (2014)

\bibitem{lawvere1966category}
Lawvere, F.W.: The category of categories as a foundation for mathematics. In:
  Proceedings of the Conference on Categorical Algebra: La Jolla 1965. pp.
  1--20. Springer (1966)

\bibitem{lawvere1997toposes}
Lawvere, F.W.: Toposes of laws of motion. American Mathematical Society,
  Transcript from video, Montreal-September  \textbf{27}, ~1997 (1997)

\bibitem{leibniz1989monadology}
Leibniz, G.W.: The Monadology: 1714 (Reprinted Edition). Springer (1989)

\bibitem{lisi2010unification}
Lisi, A.G., Smolin, L., Speziale, S.: Unification of gravity, gauge fields and
  higgs bosons. Journal of Physics A: Mathematical and Theoretical
  \textbf{43}(44),  445401 (2010)

\bibitem{loll2022quantum}
Loll, R., Fabiano, G., Frattulillo, D., Wagner, F.: Quantum gravity in 30
  questions. arXiv preprint arXiv:2206.06762  (2022)

\bibitem{loll2019quantum}
Loll, R.: Quantum gravity from causal dynamical triangulations: a review.
  Classical and Quantum Gravity  \textbf{37}(1),  013002 (2019)

\bibitem{lurie2009higher}
Lurie, J.: Higher topos theory. Princeton University Press (2009)

\bibitem{lurie2009classification}
Lurie, J.: On the classification of topological field theories. In: Current
  developments in mathematics, 2008, pp. 129--280. International Press of
  Boston (2009)

\bibitem{meschini2005geometry}
Meschini, D., Lehto, M., Piilonen, J.: Geometry, pregeometry and beyond.
  Studies in History and Philosophy of Science Part B: Studies in History and
  Philosophy of Modern Physics  \textbf{36}(3),  435--464 (2005)

\bibitem{misner1973gravitation}
Misner, C.W., Thorne, K.S., Wheeler, J.A.: Gravitation. Macmillan (1973)

\bibitem{sep-justep-coherence}
Olsson, E.: {Coherentist Theories of Epistemic Justification}. In: Zalta, E.N.
  (ed.) The {Stanford} Encyclopedia of Philosophy. Metaphysics Research Lab,
  Stanford University, {F}all 2021 edn. (2021)

\bibitem{oriti2006group}
Oriti, D.: The group field theory approach to quantum gravity. arXiv preprint
  gr-qc/0607032  (2006)

\bibitem{patton1975physics}
Patton, C.M., Wheeler, J.A.: Is physics legislated by cosmogony? Quantum
  gravity pp. 538--605 (1975)

\bibitem{Program2013}
Program, T.U.F.: Homotopy Type Theory: Univalent Foundations of Mathematics.
  Institute for Advanced Study (2013),
  \url{https://homotopytypetheory.org/book/}

\bibitem{prokopenko2019self}
Prokopenko, M., Harr{\'e}, M., Lizier, J., Boschetti, F., Peppas, P., Kauffman,
  S.: Self-referential basis of undecidable dynamics: From the liar paradox and
  the halting problem to the edge of chaos. Physics of life reviews
  \textbf{31},  134--156 (2019)

\bibitem{randall2006warped}
Randall, L.: Warped passages: Unravelling the universe's hidden dimensions.
  Penguin UK (2006)

\bibitem{elshatlawy2023ruliology}
Rickles, D., Elshatlawy, H., Arsiwalla, X.D.: Ruliology: Linking computation,
  observers and physical law. arXiv preprint arXiv:2308.16068  (2023)

\bibitem{rideout2009emergence}
Rideout, D., Wallden, P.: Emergence of spatial structure from causal sets. In:
  Journal of Physics: Conference Series. vol.~174, p. 012017. IOP Publishing
  (2009)

\bibitem{Riehl2017a}
{Riehl}, E., {Shulman}, M.: A type theory for synthetic \(\infty\)-categories.
  Higher Structures  \textbf{1}(1),  147--224 (2017)

\bibitem{riehl2018elements}
Riehl, E., Verity, D.: Elements of $\infty$-category theory. Preprint available
  at www. math. jhu. edu/\~{} eriehl/elements. pdf  (2018)

\bibitem{risi2007geometry}
Risi, V.: Geometry and Monadology: Leibniz's analysis situs and Philosophy of
  Space. Springer (2007)

\bibitem{rovelli1998strings}
Rovelli, C.: Strings, loops and others: a critical survey of the present
  approaches to quantum gravity. arXiv preprint gr-qc/9803024  (1998)

\bibitem{rovelli2008loop}
Rovelli, C.: Loop quantum gravity. Living reviews in relativity
  \textbf{11}(1),  1--69 (2008)

\bibitem{sanders2002modified}
Sanders, R.H., McGaugh, S.S.: Modified newtonian dynamics as an alternative to
  dark matter. Annual Review of Astronomy and Astrophysics  \textbf{40}(1),
  263--317 (2002)

\bibitem{Schreiber2013a}
Schreiber, U.: Differential cohomology in a cohesive infinity-topos  (Oct
  2013), \url{https://arxiv.org/abs/1310.7930}

\bibitem{schreiber2016higher}
Schreiber, U.: Higher prequantum geometry (2016)

\bibitem{Schreiber2012}
Schreiber, U., Shulman, M.: Quantum gauge field theory in cohesive homotopy
  type theory. In: Duncan, R., Panangaden, P. (eds.) Proceedings 9th Workshop
  on Quantum Physics and Logic, {QPL} 2012, Brussels, Belgium, 10-12 October
  2012. {EPTCS}, vol.~158, pp. 109--126 (2012). \doi{10.4204/EPTCS.158.8}

\bibitem{sharma2023assembly}
Sharma, A., Cz{\'e}gel, D., Lachmann, M., Kempes, C.P., Walker, S.I., Cronin,
  L.: Assembly theory explains and quantifies selection and evolution. Nature
  pp.~1--8 (2023)

\bibitem{Shulman2016}
Shulman, M.: Homotopy type theory: A synthetic approach to higher equalities
  (Jan 2016). \doi{10.1093/oso/9780198748991.003.0003}

\bibitem{Shulman2017}
Shulman, M.: Homotopy type theory: the logic of space  (Mar 2017).
  \doi{10.1017/9781108854429.009}

\bibitem{sep-epistemology}
Steup, M., Neta, R.: {Epistemology}. In: Zalta, E.N. (ed.) The {Stanford}
  Encyclopedia of Philosophy. Metaphysics Research Lab, Stanford University,
  {F}all 2020 edn. (2020)

\bibitem{swingle2018spacetime}
Swingle, B.: Spacetime from entanglement. Annual Review of Condensed Matter
  Physics  \textbf{9},  345--358 (2018)

\bibitem{verlinde2017emergent}
Verlinde, E.: Emergent gravity and the dark universe. SciPost Physics
  \textbf{2}(3), ~016 (2017)

\bibitem{weinberg1989cosmological}
Weinberg, S.: The cosmological constant problem. Reviews of modern physics
  \textbf{61}(1), ~1 (1989)

\bibitem{wheeler1980pregeometry}
Wheeler, J.A.: Pregeometry: Motivations and prospects  (1980)

\bibitem{wigner1960unreasonable}
Wigner, E.P.: The unreasonable effectiveness of mat hematics in the natural
  sciences. Commun. Pure Appl. Math  \textbf{XIII},  1--14 (1960)

\bibitem{Wolfram2002a}
Wolfram, S.: A new kind of science. Wolfram Media, Champaign, Ill. (2002)

\bibitem{Wolfram2020}
Wolfram, S.: A class of models with the potential to represent fundamental
  physics. Complex Systems  \textbf{29}(2) (2020).
  \doi{10.25088/complexsystems.29.2.107},
  \url{https://www.complex-systems.com/abstracts/v29_i02_a01/}

\bibitem{wolframruliad}
Wolfram, S.: The concept of the ruliad  (2021),
  \url{https://writings.stephenwolfram.com/2021/11/the-concept-of-the-ruliad/}

\bibitem{wolframconc}
Wolfram, S.: What is consciousness? some new perspectives from our physics
  project  (2021),
  \url{https://writings.stephenwolfram.com/2021/03/what-is-consciousness-some-new-perspectives-from-our-physics-project/}

\bibitem{wolfram2022metamathematics}
Wolfram, S.: Metamathematics: Foundations \& Physicalization. Wolfram Media
  (2022)

\bibitem{zapata2022invitation}
Zapata-Carratala, C., Arsiwalla, X.D.: An invitation to higher arity science.
  arXiv preprint arXiv:2201.09738  (2022)

\bibitem{zapata2022heaps}
Zapata-Carratala, C., Arsiwalla, X.D., Beynon, T.: Heaps of fish: arrays,
  generalized associativity and heapoids. arXiv preprint arXiv:2205.05456
  (2022)

\bibitem{zapata2023beyond}
Zapata-Carratal{\'a}, C., Schich, M., Beynon, T., Arsiwalla, X.D.: Hypermatrix
  algebra and irreducible arity in higher-order systems: Concepts and
  perspectives. Advances in Complex Systems  (2023)

\end{thebibliography}
\end{document}